\documentclass{elsart}

\usepackage{psfig}

\usepackage{natbib}

\begin{document}

\runauthor{Elvis}


\begin{frontmatter}

\title{A Structure for Quasars}

\author[CfA]{Martin Elvis}
\address[CfA]{Harvard-Smithsonian Center for Astrophysics
Cambridge MA 01238 USA}

\begin{abstract}
We propose a simple unifying structure for the inner regions of
quasars and AGN. This empirically derived model links together
the BALs, the narrow UV/X-ray ionized absorbers, the BELR, and
the Compton scattering/fluorescing regions into a single
structure. The model also suggests an origin for the
large-scale bi-conical outflows. Some other potential
implications of this structure are discussed.

\end{abstract}


\end{frontmatter}


\section{Broad Absorption Line and Warm Absorber Unification}

The starting point lies with the narrow absorption
lines (NALs)/X-ray warm absorbers in the well-studied Seyfert 1
galaxy NGC~5548. Accepting for a moment the identity of the UV
and X-ray absorbers, then the absorber must lie $\sim$10$^{16}$~cm
from the continuum source, but is only $\sim$10$^{15}$~cm
thick; i.e. a thin shell. (This is also needed to have a single
ionization state fit the data.) This shell has an outward
velocity of 1200~km~s$^{-1}$ (figure~1), yet in 20 years has not
changed its ionization state. (Being a trace ion, N(CIV) should
have increased by a factor 100, unless density changes precisely
matched the declining ionizing continuum, yet the CIV EW has been
constant.)  Mathur, Elvis \& Wilkes (1995) explained this
constancy by invoking a flow of absorbing material across our
line of sight to the continuum. So the structure remained
constant, although the particular gas atoms changed.

A flow across the line of sight implies a conical structure
(figure~1c), so from some directions no absorbers will be
seen. A cone angle of 60$^{\circ}$ divides the solid angle
equally, and could explain why half of Seyfert galaxies have
ionized absorbers and half do not (figures~3a,b).

\begin{figure}[htb]
\centerline{\psfig{figure=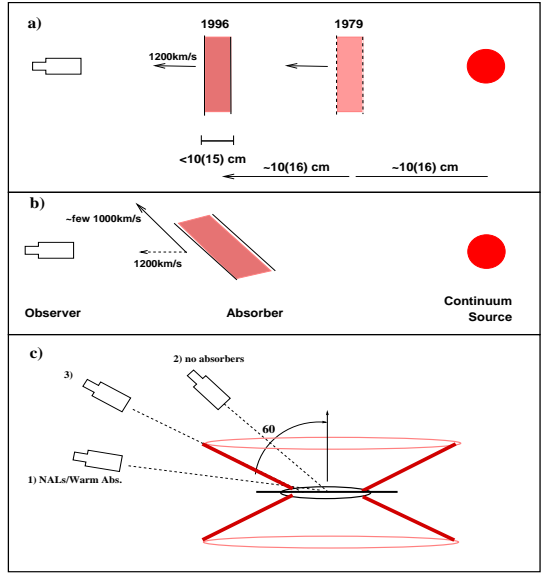,height=3.0truein,width=2.8truein,angle=0}}
\caption{Geometry of the absorber in NGC~5548: (a) the absorber
is thin and has a velocity of 1200~km~s$^{-1}$ out from the
nucleus; (b) to sustain constant ionization over 20 years
requires a flow across the line-of-sight; (c) a bi-conical
geometry is the simplest structure that allows (b).}
\end{figure}

There is a third viewing direction: directly down the flow.  This
too will show a highly ionized absorber, but with much larger
column density and higher velocity. Recent work on broad
absorption line (BAL) quasars suggests that they have such
conditions (e.g. Mathur, Elvis \& Singh 1996; Hamman 1998). If the
flow opens at a half angle of 6$^{\circ}$ then 10\% of AGN will
be seen as BAL quasars, as observed. This is appealing since both
warm absorber outflows (see above) and BALs are likely present in
all quasars (to explain, among other reasons, the strong
continuum polarization in BAL troughs, Ogle 1998).

However the velocities in BALs are $\sim$10 times those of the
NALs. The simplest modification to figure~1 that accomodates this
is to bend the flow (figure~2, figure~3a). At first glance this
geometry appears unlikely; however, it can arise quite naturally.
Symmetry requires that a disk wind will first rise vertically.
Once exposed to radiation pressure from the continuum source the
flow will begin to bend outward, giving the
$\sim$1000~km~s$^{-1}$ NAL outflow velocity, and will become
radial when $v(vertical)=v(radial)$. The `detachment velocities'
of BALs measure this critical velocity to be 0-5000~km~s$^{-1}$,
so $v(vertical)\sim v(Keplerian)$, the disk orbital velocity at
the wind ejection point.

\begin{figure}[htb]
\centerline{\psfig{figure=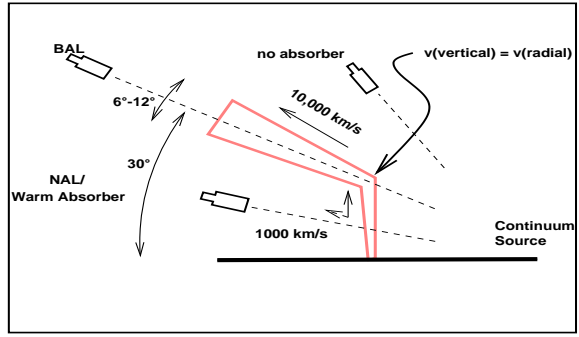,height=1.75truein,width=3.0truein,angle=0}}
\caption{Simplest geometry needed to unite BAL
and NAL (UV/X-ray) absorbers.}
\end{figure}

\section{WHIM Outflow and BELR Unification}

Conditions in the `warm' absorber are likely to be high density
($n_e\ge$10$^7$cm$^{-3}$), and truly warm ($\sim$10$^6$K,
Nicastro et al., 1999, 2000), hence the name `Warm Highly Ionized
Medium' (WHIM). The WHIM pressure is comparable with that in
broad emission-line clouds.  The geometry removes the objection
that a BELR confining medium would be Compton thick, while the
bulk co-outflowing of both WHIM and BELR means that the BELR
clouds will not be destroyed by shear forces. In addition:
$v(vertical) \sim v(BELR)$ and $r(WHIM)\sim r(BELR, CIV)$, about
10$^{16}$cm in NGC~5548. The cylinder/cone geometry of the wind
naturally divides the BELR into a high ionization (cylindrical)
region, and a low ionization (conical) region shielded from
X-rays by the WHIM.

\begin{figure}[htb]
\centerline{\psfig{figure=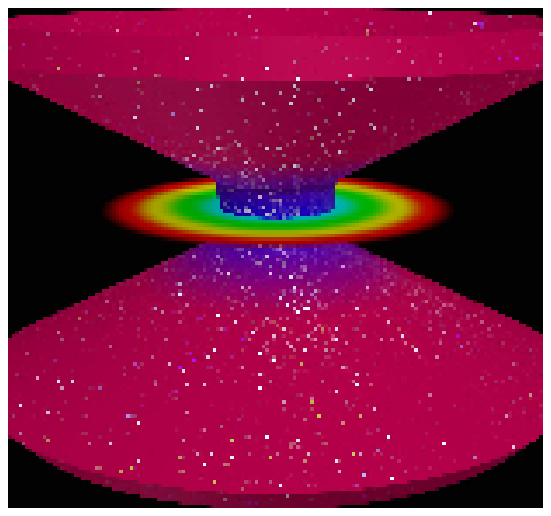,height=2.0truein,width=4.0truein,angle=0}}
\centerline{\psfig{figure=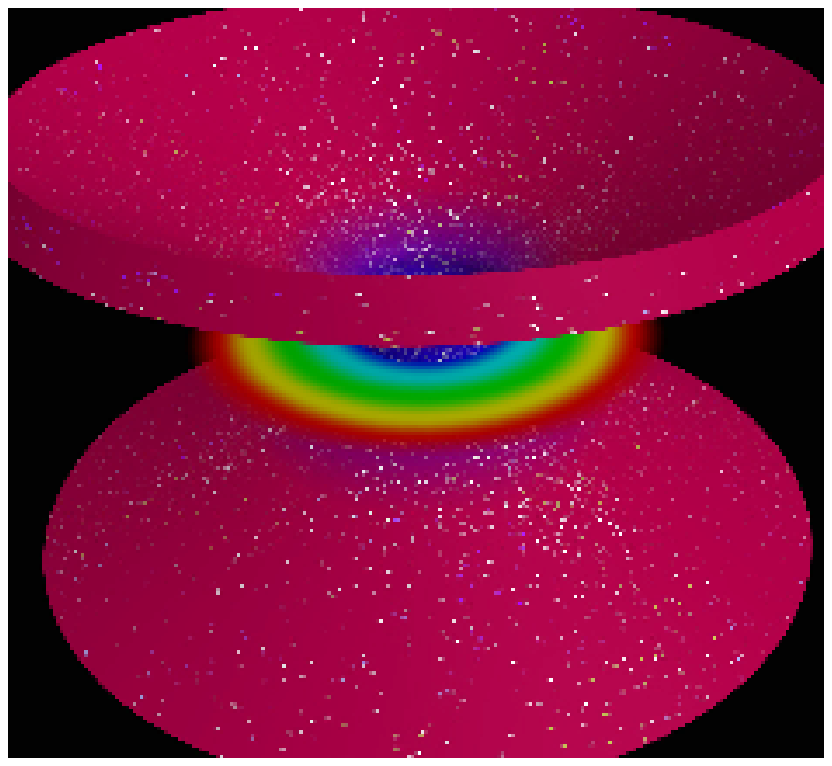,height=2.0truein,width=4.0truein,angle=0}}
\centerline{\psfig{figure=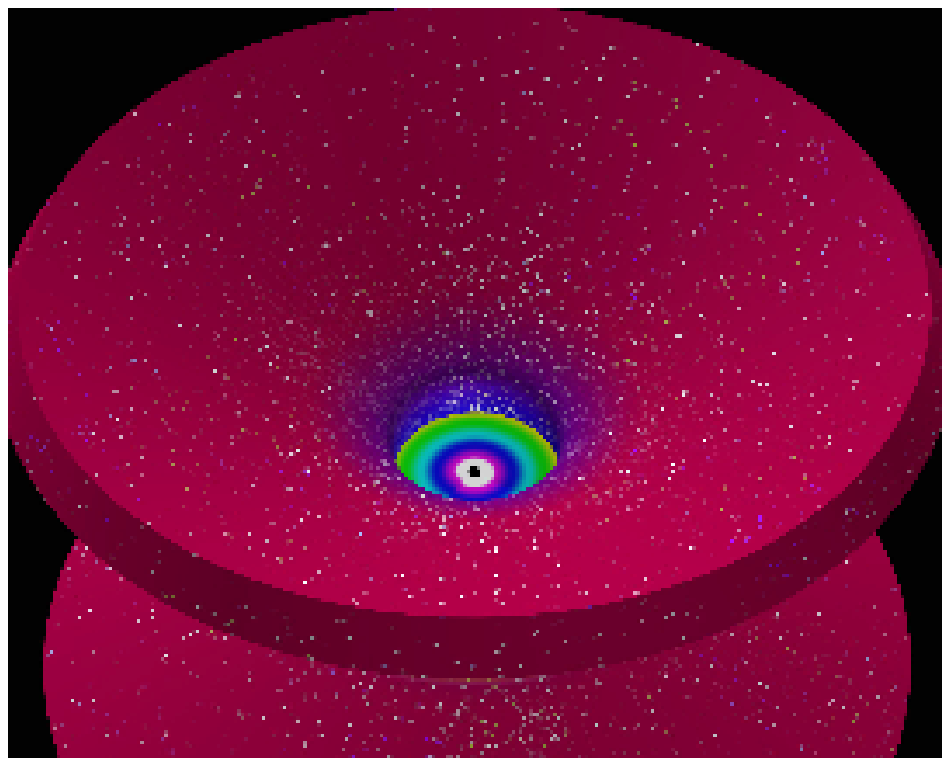,height=2.0truein,width=4.0truein,angle=0}}
\caption{Side (NAL, top panel), along flow (BAL, middle panel) and top
(bottom panel) views of the proposed quasar structure. [{\small\em
Sharp edges are an artifact of the rendering program.}]  }
\end{figure}

\section{WHIM Outflow and Scattering/Fluorescence Sites}

Along the flow (in the BAL direction) the WHIM is Compton
thick. Scattering from the rear of the flow (figure~3c) produces
the polarized continuum in BALs (Ogle 1998).  Compton thick
scatterers are also needed to make: `polarized BEL' Seyfert~2s
(e.g. NGC~1068); UV continuum polarization below Ly$\alpha$
(Beloborodov \& Poutanen 1999); and the X-ray reflection
features, the `Compton hump' and fluorescent Fe-K line.  Could
these could arise from the WHIM?

The UV continuum polarization needs semi-relativistic outflow,
which the conical BAL part of the wind has. The `polarized BEL'
might work if there is dust in the outer part of the wind. (If
this happens only in low luminosity AGN, then the missing
`quasar~2s' are the BAL quasars.) However HST has resolved some
of the `mirrors' in Seyfert~2s on $\sim$10~parsec scales, too
large to be the WHIM. The WHIM will necessarily produce X-ray
features, proportional to the WHIM covering factor of 10\%--
30\%, i.e. almost half the observed Fe-K line strength. The WHIM
Fe-K component cannot vary on timescales $<$ weeks.

\section{Extensions: The Baldwin effect; Bi-cones; Torus}

High luminosity quasars differ from the lower luminosity AGN in
several features. Changes in the outflow opening angle and the
height of the cylindrical region, e.g. from increased radiation
or cosmic ray pressure at high luminosities, may explain: the
lower CIV EW (Baldwin effect), the rarity of NALs, and the
weakness of X-ray scattering features in high luminosity quasars.
The need for an `obscuring torus' is weaker in this model: (1)
the Seyfert~2s with polarized BELs could be viewed through a
dusty BAL; (2) the conical shell outflow could itself produce the
bi-cone structures, which would then be hollow, matter bounded,
cones rather than filled, ionization bounded, structures; (3) the
more common `no polarized BEL' Seyfert~2s could be made in
several ways (Lawrence \& Elvis 1982).

\begin{figure}[htb]
\centerline{\psfig{figure=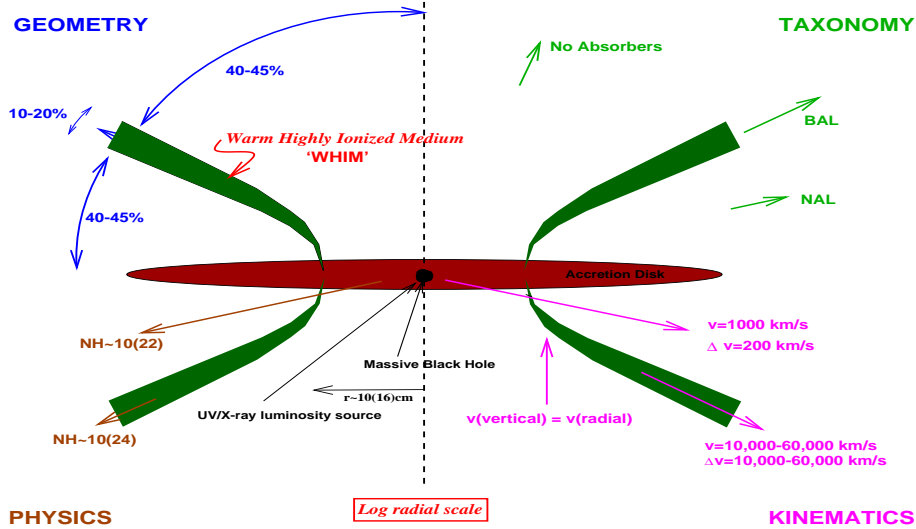,height=2.75truein,width=4.75truein,angle=-90}}
\caption{The proposed structure for quasars.  The four quadrants
illustrate (clockwise from top left): the covering factors; the
spectroscopic appearance to a distant observer at various angles;
the outflow velocities along different lines of sight; some
representative radii (scaled to the Seyfert 1 galaxy NGC~5548)
and column densities.}
\end{figure}

\section{Summary}

The complete structure proposed here and in Elvis (2000) is shown
in figure~4. This ambitious proposal draws together a number of
previously disparate areas of quasar research into a single
simple scheme. The broad and narrow absorption line (BAL, NAL)
regions, the high and low ionization broad emission line regions
(BELR), and at least some of the five Compton thick scattering
regions can all be combined into the single funnel-shaped
outflow, while on large scales this outflow could produce the
bi-conical narrow emission line regions. This unification gives
the model a certain appeal.



\end{document}